\documentclass[11pt]{article}
\usepackage{cite}
\usepackage{amsmath,amsfonts,amssymb}
\usepackage[small,bf,hang]{caption}
\usepackage{slashed}
\usepackage{mathabx}
\usepackage{latexsym,epsfig}
\usepackage{color}
\usepackage{mathtools}

\def\hybrid{
        \topmargin -20pt
        \oddsidemargin 0pt
        \headheight 0pt \headsep 0pt
        \textwidth 6.25in 
        \textheight 9.5in 
        \marginparwidth .875in
        \parskip 5pt plus 1pt \jot = 1.5ex}

\hybrid

 \csname
@addtoreset\endcsname{equation}{section}

\def\moth{\mathsurround=0pt}
\newdimen\zo \zo=0pt

\def\tick{\leaders\hrule height 0.5ex depth 0pt \hskip 0.5pt}
\def\upboxfill{$\moth \setbox\zo\hbox{\tick}%
  \hskip 3pt\hbox to 0pt{$\tick$\hss}\hrulefill \hbox to 7.5pt{$\tick$\hss}$}

\def\dtick{\leaders\hrule height .34pt depth 0.5ex \hskip 0.5pt}
\def\downboxfill{$\moth \setbox\zo\hbox{\dtick}%
  \hskip 2pt\hbox to 0pt{$\dtick$\hss}\hrulefill \hbox to 2pt{$\dtick$\hss}$}

\def\nl{\nonumber\\}
\def\bec{\begin{center}}
\def\ec{\end{center}}
\def\a{\alpha}  

\def\b{\beta}  
\def\c{\gamma} 

\def\d{\delta} 
\def\D{\Delta}

\def\l{\lambda}

\def\r{\rho}
\def\s{\sigma}

\def\t{\tau}

\def\pa{{\partial }}
\def\Ra{{\Rightarrow}}

\def\bfR{{\bf{R}}}
\def\Asl{{A\!\!\!/}}
\def\pasl{{\partial\!\!/}}
\def\bE{{\bf{1}}}
\def\bX{{\bf{X}}}
\def\bY{{\bf{Y}}}

\def\cD{{\cal D}}

\def\cO{{\cal O}}

\def\cR{{\cal R}}
\def\cS{{\cal S}}

\def\cO{{\cal O}}

\def\nn{\nonumber}

\def\Tr{{\rm Tr}\,}

 \def\det{{\rm det\,}}
\def\be{\begin{equation}}
\def\ee{\end{equation}}
\def\bea{\begin{eqnarray}}
\def\eea{\end{eqnarray}}
\def\ba{\begin{array}}
\def\ea{\end{array}}

\begin{document}

\begin{titlepage}
\begin{flushright}    
{\small $\,$}
\end{flushright}
\vskip 1cm
\centerline{\LARGE{\bf{Supersymmetric Yang-Mills Theories:}}}
\vskip 0.5cm
\centerline{\LARGE{\bf{not quite the usual perspective}}}
\vskip 1.5cm
\centerline{Sudarshan Ananth$^{\,\dagger}$, Hermann Nicolai$^{\,\star}$, Chetan Pandey$^{\,\dagger}$ and Saurabh Pant$^{\,\dagger}$}
\vskip 0.5cm
\centerline{${\,}^\dagger$\it {Indian Institute of Science Education and Research}}
\centerline{\it {Pune 411008, India}}
\vskip 0.5cm
\centerline{${\,}^*$\it {Max-Planck-Institut f\"ur Gravitationsphysik (Albert-Einstein-Institut)}}
\centerline {\it {Am M\"{u}hlenberg 1, 14476 Potsdam, Germany}}
\vskip 1.5cm
\centerline{\bf {Abstract}}
\vskip .5cm
In this paper, we take up an old thread of development concerning the characterization
of supersymmetric theories without any use of anticommuting variables that goes back to one of 
the authors' very early work \cite{Nic1}. Our special focus here will be on the formulation of 
supersymmetric Yang-Mills theories, extending previous results beyond $D=4$ dimensions.  
This perspective is likely to provide new insights into these theories, 
and in particular the maximally extended $N=4$ theory. As a new result we re-derive the admissible 
dimensions for interacting (pure) super-Yang-Mills theories to exist. 
\newline This article is dedicated to the memory of Peter Freund, amongst many other things an early contributor to supersymmetry, and an author of one of the very first papers on superconformal gauge theories \cite{Freund1}. The final section contains some personal reminiscences of H.N.'s encounters with Peter Freund.  
\vfill
\end{titlepage}



\renewcommand\footnoterule{}

\thispagestyle{empty}
\mbox{ }
\vspace{5mm}

\renewcommand{\thefootnote}{\fnsymbol{footnote}}

\section{Introduction and Conventions}

There is a large literature on supersymmetric Yang-Mills theories (see e.g. \cite{SYM0}), particularly concerning the maximally extended $N=4$ theory in four dimensions, or equivalently,  pure super-Yang-Mills theory in $D=10$ \cite{SYM1}. This theory has been proposed to underlie M theory, either via the AdS/CFT correspondence \cite{Mal}, or, in its dimensionally reduced form, the maximally supersymmetric SU($\infty$) matrix model \cite{Matrix1,Matrix2}. In view of these far reaching conjectures it appears worthwhile and expedient to investigate supersymmetric Yang-Mills theories and their properties from {\em every} possible angle\footnote{In particular to confirm the often heard claim that maximally extended $N=4$ theory {\em defines} non-perturbative quantum gravity in the AdS$_5$ bulk in terms of the boundary theory via a holographic correspondence. At the very least, a full validation of this statement would require a {\em non-perturbative} definition of the boundary theory itself, which is not yet available.}.
Here we attempt to do so by following a route  different from the one usually taken. This goes back to the early work of one of the authors \cite{Nic1,Nic2} which has been followed up on only intermittently since the mid-eighties, after important early work by Flume and Lechtenfeld \cite{FL},  Dietz and Lechtenfeld \cite{DL1,DL2}, and an intriguing attempt at a closed form expression for the half-maximal $N=2$ theory by de Alfaro, Fubini, Furlan and Veneziano \cite{Fub1}. 
In this paper we extend, in a minor way, the old results of \cite{Nic1} by showing that the constructions presented there extend to all pure supersymmetric Yang-Mills theories and, in particular the maximally extended $N=4$ theory. We also clarify the link between the results of \cite{Nic1,Nic2} and \cite{FL,DL1,DL2,L1}. New results presented here are part of a larger ongoing project \cite{SYM2} where amongst other things, we shall extend these considerations to the next order in the coupling constant. In view of the huge body of results on the $N=4 $ theory, obtained mostly in the context of the AdS/CFT correspondence, there are numerous directions to be explored.

The main result of \cite{Nic1} can be summarized as follows: for any rigidly supersymmetric theory with at most quadratic fermionic terms in the Lagrangian, there exists a non-linear and non-local transformation of the bosonic fields (``Nicolai map") that linearizes the bosonic action in such a way that the Jacobian of the bosonic field transformation equals the determinant (Berezinian) obtained upon integrating out all anticommuting fields. Specializing right away to supersymmetric gauge theories (the case of interest here), the statement is that the gauge fields admit a non-linear and non-local transformation
\be
T_g[A]_\mu^a(x) \equiv A^{'a}_\mu(x,g;A)
\ee
(where $g$ is the Yang Mills coupling constant) with the following properties:

\begin{enumerate}
\item Substitution of $A'(A)$ into the free Maxwell action (or rather: sum of Maxwell actions)
yields the interacting theory, {\em viz.}
\be
\cS_0[A'(A)] = \cS_g[A] \equiv \frac14 \int d^D x\,  F_{\mu\nu}^a F_{\mu\nu}^a
\ee 
where
\be\label{Fmn}
F_{\mu\nu}^a \,\equiv \,  \pa_\mu A_\nu^a - \pa_\nu A_\mu^a + gf^{abc} A_\mu^b A_\nu^c
\ee
is the Yang-Mills field strength [with fully antisymmetric structure constants $f^{abc}$
for the chosen gauge group, usually SU($n$)], and $\cS_0$ is the free Maxwell action 
(that is, $\cS_g$ for $g=0$). The statement also remains correct with a gauge fixing term,
cf. (\ref{gf2}) below.
\item The Jacobian of the transformation equals the product of the Matthews-Salam-Seiler 
(MSS) determinant (or Pfaffian) \cite{MSS} obtained by integrating out the gauginos and the 
Faddeev-Popov (FP) determinant \cite{FP} (obtained by integrating out the ghost fields $C^a,\bar{C}^a$), 
\be\label{Det}
\det \left(\frac{\delta A^{'a}_\mu(x,g;A)}{\delta A^b_\nu(y)}\right) = \Delta_{MSS} [A]\;\Delta_{FP}[A]
\ee
at least in the sense of formal power series.
\end{enumerate}

One can thus characterize an important class of rigidly supersymmetric theories in a way that makes no use of anticommuting variables at all. In this contribution we will explain this result (which for $D=4$ super-Yang-Mills theory was obtained and proved long ago in\cite{Nic1,Nic2,FL,DL1,DL2}) in simple terms by explicitly rederiving the map up to $\cO(g^2)$, and extending it to all pure supersymmetric Yang-Mills theories (analogous results also hold for 
non-anomalous matter coupled  supersymmetric  gauge theories, but these will be of no concern here). 
As a new result, using this approach we will recover the well-known result of \cite{SYM1} that interacting
pure supersymmetric Yang-Mills theories can exist only in space-time dimensions $D=3,4,6,10$
(for the free theories this is simply a consequence of the equality of bosonic and fermionic 
degrees of freedom on shell).
At least as far as the known results are concerned, this formalism does not care about the question of whether there exists an off-shell formulation, and could in principle even provide a (non-perturbative) regulator of these theories that could preserve basic features of supersymmetry and gauge invariance even in the regulated theory, though in disguised form. Accordingly, this approach adopts the opposite strategy from the usual one, of introducing ever more auxiliary and ghost degrees of freedom which in turn must be removed by yet more auxiliary gauge transformations, with commuting and anticommuting parameters, involving superspace formulations, Wess-Zumino gauges, and the like.

We start with some conventions. We will be slightly cavalier about the space-time signature, which can be taken to be either Euclidean (as in \cite{Nic1,Nic2}) or Minkowskian (as in \cite{FL,DL1,DL2}). The Minkowskian signature is perhaps more convenient if one wants to avoid issues concerning the existence (or not) of Majorana spinors in Euclidean space-times. The usual assumption that (interacting) functional measures have a better chance of being rigorously defined when using a Euclidean signature is actually not so relevant in view of the fact that {\em Gaussian} functional measures are well-defined even with imaginary (oscillatory) exponents, via their 2-point correlators and Wick's theorem. Ideally, this is all that is needed here --- of course, provided one can succeed in producing a closed form expression for the map $T_g$, or something close to it, which is no small order! Such closed form solutions indeed exist for special models, such as supersymmetric quantum mechanics \cite{CG}, as well as certain Wess-Zumino-type or Landau-Ginzburg-type $N=2$ models in two dimensions (see \cite{L2}, and \cite{KK} for recent results). Alternatively, one 
can simply regard the main formulas in section~3 as analytic continuations of the corresponding Minkowskian ones, even independently of their derivation.

We will need covariant derivatives only for the adjoint representation (in which the 
gauginos also transform); they are
\be \label{Dmu}
D_\mu V^a \,\equiv \, \pa_\mu V^a + g f^{abc} A_\mu^b V^c  \;\; \Rightarrow \quad
[D_\mu , D_\nu] \, V^a = g f^{abc} F_{\mu\nu}^b V^c
\ee 
with the Yang-Mills field strength (\ref{Fmn}).

The free scalar propagator is (with the Laplacian $\Box \equiv \pa^\mu \pa_\mu$)
\be
C(x) = \int \frac{d^D k}{(2\pi)^D} \frac{e^{ikx}}{k^2} \quad \Ra \quad
- \Box C(x) = \d(x) 
\ee
where $\d(x) \equiv \d^{(D)}(x)$ is the $D$-dimensional $\d$-function. 
For arbitrary $D$ we have
\be
C(x) = \frac1{(D-2)D\pi^{D/2}} \, \Gamma\left( \frac{D}2 +1\right) (x^2)^{1- \frac{D}2} \; ;
\ee
in particular, for $D=4$
\be
C(x) = \frac1{4\pi^2}\cdot \frac1{x^2}
\ee
When writing $\pa_\l C(x-y) \equiv (\pa/\pa x^\l) C(x-y) \equiv \pa_\l^x C(x-y)$, the derivative by convention {\em always acts on the first argument}. Careful track needs to be kept of the sign flips $\pa_\l^x C(x-y) = - \pa_\l^y C(x-y) = + \pa_\l^x C(y-x) = - \pa^y_\l C(y-x)$,

The free fermionic propagator is 
\be
\c^\mu\pa_\mu S_0(x)   = \d (x)   \quad \Ra \quad S_0 (x) = - \c^\mu \pa_\mu C(x)
\ee
where the spinor indices are suppressed. This implies $S_0(x-y) = - S_0(y-x)$. The effective number of fermionic degrees of freedom (spinor components) will be designated by $r_D$, and of course depends on $D$ including extra factors of $\frac12$ for Majorana or Weyl spinors, and $\frac14$ for Majorana-Weyl spinors, respectively. For pure supersymmetric Yang-Mills theories the only possibilities are
\be\label{Dr}
D\,=\,  3,4,6,10  \qquad \Longleftrightarrow \qquad r_D\,=\,  2,4,8,16
\ee
With Minkowskian signature, for $D=4$ space-time this corresponds to a Majorana spinor, for $D=6$ to a Weyl spinor, while for $D=10$ we have one more factor of $\frac12$ because of the  Majorana-Weyl condition. We shall rederive this constraint in section~3 {\em without} any use of anticommuting objects.

We also need the fermionic propagator in a gauge-field dependent background characterized by $A_\mu^a(x)$
\be
\c^\mu (D_\mu S)^{ab}(x) \equiv \c^\mu \Big[ \d^{ac}\pa_\mu - gf^{acd} A_\mu^d(x) \Big] S^{cb}(x)
     = \d^{ab} \d(x)
\ee 
Using the standard relation $(1-X)^{-1} = 1 + X + X^2 + \cdots$ the full propagator can be expanded in terms of free propagators and the background gauge field as
\be\label{S}
S^{ab}(x,y;A) = S_0^{ab}(x-y) \,+\, g \int du  \, S_0^{ac}(x-u) f^{cdm} \Asl^m(u) S_0^{db}(u-y) 
\,+\, \cdots 
\ee
Below we will also use the shorthand notation
\be
S^{ab} = S_0^{ab} + g (S_0 * \Asl * S_0)^{ab} + g^2(S_0 * \Asl * S_0 *\Asl * S_0)^{ab} \, + \cdots
\ee
for such expansions, with the convention that the contraction of the structure constant with the gauge field is usually through the last index, as displayed above.

Although the formalism works for other gauge choices as well\footnote{In particular the axial, and more specifically, the light-cone gauge, which is of special interest in view of a possible link with the results of \cite{ABR}, \cite{AT}.}, we  will consider only one 
gauge fixing function here, namely the Landau gauge
\be
G^a[A_\mu] = \pa^\mu A_\mu^a 
\ee
The functional integral over gauge fields will thus be understood to contain a 
$\delta$-functional implementing the gauge condition, that is
\be\label{deltafunctional}
\int \cD A_\mu^a\,  (\cdots )\;\; \rightarrow \; \int \cD[A_\mu^a] \, \prod_{x, a}
\delta\Big(\partial^\mu A^a_\mu(x)\Big) \, (\cdots )
\ee
The ghost propagator 
\be
G^{ab} (x) \equiv \underbracket[0.5pt]{C^a(x) \bar{C}^b}(0)
\ee
obeys
\be
- \pa^\mu (D_\mu G)^{ab}(x) = \d^{ab}\d (x) 
\ee
for the Landau gauge. As with the fermions, we can expand it in terms of free propagators. While $G^{ab}(x)$ does depend on $g$ and the background field $A_\mu^a(x)$, this dependence drops out in $D_\mu G^{ab}(x)$; more specifically, we have
\be\label{DGab}
D_\mu G^{ab}(x) =  \d^{ab} \pa_\mu C(x) 
\ee
with the free propagator $C(x)$.

\section{$\cR$ prescription (Landau gauge)}

A systematic order by order construction of the {\em inverse} transformation $T_g^{-1}$, and in fact a proof of the main theorem above at least for the $N=1$, $D=4$ theory is provided by the $\cR$ prescription introduced in \cite{FL,DL1,Nic2,DL2,L1}. 
To this aim we define the $\cR$ operator 
\be \label{cR}
\cR \,\equiv\,  \frac{d}{dg} \,+\,  \bfR 
\ee
which can be viewed as the Lie algebra generator of the {\em inverse} map $T_g^{-1}$
\be\label{Tginv}
(T_g^{-1}A)_\mu^a(x) \,=\, A_\mu^a(x) \,+ \,
\sum_{n=1}^\infty \frac1{n!}  \, g^n \, \Big(\cR^n  \big[ A \big]_\mu^a(x)\Big)_{g=0}
\ee
For the Landau gauge the second part of the $\cR$ operator, $\bfR$ on $A_\mu^a$ is defined by 
\be
\label{bR1}
\bfR[A]_\mu^a(x) \,\equiv\, - \, \frac1{2r_D} \int du dv\, 
\Pi_{\mu\nu}(x-u){\rm Tr}\, \big( \c_\nu \c^{\r\s} S^{ba}(v-u)\big)
      f^{bcd} A_\r^c(v)A_\s^d(v)
\ee
with the transversal projector
\be\label{Pi1}
\Pi_{\mu\nu}(x-y) \,\equiv \, \left( \d_{\mu\nu} - \frac{\pa_\mu \pa_\nu}{\Box}\right)\d(x-y)
\, \cong \, \d_{\mu\nu} \d(x-y) + \pa_\mu C(x-y) \pa_\nu 
\ee
where "$\cong$" means equality in the sense of distributions.
\newline In  (\ref {Tginv}), we need to keep the full $g$-dependence when successively acting with $\cR$ at each step of the iteration and only set $g=0$ at the very end, before inserting the result into the Taylor series expansion (if not implemented properly, crucial contributions will be missed from $\cO(g^3)$ onwards, as outlined below). In the final step this series expansion has to be inverted to obtain the map $T_g$. This will be illustrated by the explicit calculation in the next section, where for simplicity we spell out all the relevant steps in detail (but only for the Landau gauge).

We stress that the fermionic propagator $S$ in (\ref{bR1}) is the {\em full} propagator, and hence still depends on $g$ and the background gauge field. Furthermore, this formula is valid in all relevant dimensions, with $r_D$ and $D$ related as in (\ref{Dr}). The expression in (\ref{bR1}) follows directly from the formula (4.20) in \cite{Nic2} (originally due to \cite{FL})
\be
\cR[X] \,=\, \frac{dX}{dg} \,+\, \underbracket{\d_\a X \D_\a}
\,+\,  \int \underbracket{\bar C^a \underbracket{ \d_\a G^a[A_\mu] \D_\a} s(X)}
\ee
by working out the contractions and by substituting (\ref{DGab}). The correct prefactors and signs in (\ref{bR1}) were obtained by simply comparing this formula with the first order result in \cite{Nic1} (equation (3.24)). The $\cR$ operator acts distributively, 
\be
\cR\big[A^a_\mu(x) A_\nu^b(y) \cdots \big] \, \,= \cR\big[A_\mu^a(x)\big] A_\nu^b(y)\cdots \, + \, 
    A_\mu^a(x) \cR\big[A_\nu^b(y)\big]\cdots \, + \, \dots
\ee
From (\ref{bR1}) it follows immediately that the $\cR$ operation preserves the gauge fixing function 
\be\label{gf1}
\pa^\mu \, \cR\big[ A_\mu^a(x)\big] = 0 
\ee
This will guarantee that the equality
\be\label{gf2}
\pa^\mu (T_g(A)_\mu^a) (x) = \pa^\mu A_\mu^a(x)
\ee
holds for all values of the Yang-Mills coupling constant $g$. Equations (\ref{cR}) and (\ref{bR1}) are our basic formulas, as their iterative application will yield the expansion coefficients of $T_g^{-1}$ to any desired order, though with a rapidly 
increasing number of terms.

The $\cR$ operation is compactly represented by the functional differential operator
\be
\cR = \frac{d}{dg}  - \, \frac1{2r_D} \int dx\,du\,dv\; 
\Pi_{\mu\nu}(x-u){\rm Tr}\, \big( \c_\nu \c^{\r\s} S^{ba}(v-u)\big) f^{bcd} A_\r^c(v)A_\s^d(v)
        \frac{\d}{\d A_\mu^a(x)}
\ee
In particular, it acts as follows on the full fermionic propagator
\bea
\cR\, S^{ab}(x,y;A) &=& \int du \, S^{ac}(x,u;A) f^{cdm} \c^\l A_\l^m (u) S^{db}(u,y;A)  \, - \\
    &&  \hspace{-1cm}  -\, \frac{g}{2r_D} \int du\, dv\, dw\; 
\Pi_{\mu\nu}(w-u){\rm Tr}\, \big( \c_\nu \c^{\r\s} S^{pe}(v-u)\big) f^{pmn} A_\r^m(v)A_\s^n(v)
     \, \times \nn\\
   &&     \hspace{2cm} \times \;  S^{ac}(x,w;A) f^{cde} \c_\mu S^{db}(w,y;A)                   \nn
\eea
Importantly, the second term comes with a factor of $g$ and will therefore drop out upon setting $g=0$. However, it will contribute at the next order when acting again with $d/dg$ and then setting $g=0$; this extra contribution will appear from the third order onwards. As mentioned already, the above prescription generates the {\em inverse} map, and can, in principle,  be used to calculate $T_g^{-1}$ to arbitrary order. However, while $\cO(g^2)$  is still fairly straightforward to work out as shown below, the procedure quickly  becomes complicated at higher orders and is already cumbersome to evaluate at $\cO(g^3)$ \cite{SYM2}.\footnote{The results for $T_g^{-1}$  up to $\cO(g^3)$ can already been found in \cite{DL2}, but only in an implicit form, where the $\c$-traces have not been evaluated, and $T^{-1}_g$ has not been inverted to determine the map $T_g$ itself up to this order.} 

Independent of the question of whether the series expansion (\ref{Tginv}) and its inverse (in the sense of a formal power series) can be
elevated to closed form expressions, it is a remarkable feature that these series admit a {\em finite} radius of convergence (with suitable norms on the function space of gauge field backgrounds). This follows by inspection of the $\cR$-operation, which can be seen to generate $\cO(c \,n)$ new terms at the $n$-th step of the iteration, and hence only $\cO(c^n \,n! )$ terms at the $n$-th order $\cO(g^n)$ (where $c$ is a model dependent constant). The well known
combinatorial divergences of the quantized theory, with extra factors of $n!$, are then generated upon quantization in terms of the {\em free} field $A^{'a}_\mu$, and more specifically after contracting gauge field lines in the tree-like expansion of $T_g^{-1}$ in all possible ways \cite{Nic2,DL1}.

\section{Lowest order computations to $\cO(g^2)$} 
For the Landau gauge, the lowest order result is obtained by simply setting $g=0$ in (\ref{bR1}) (the $\frac{d}{dg}$ piece being trivially zero at lowest order)
\bea\label{Og1}
\cR\big[ A\big]_\mu^a (x) \Big|_{g=0}   \,&=&\,
- \, \frac1{2r_D} \int du\, {\rm Tr}\, \big( \c_\mu \c^{\r\s} S_0^{ba}(u-x)\big) f^{bcd} A_\r^c(u)A_\s^d(u) \nl[2mm]
 &=&\, -\,  \int du\, \pa_\l C(x-u) f^{abc} A_\mu^b(u) A_\l^c(u)
\eea
The ghost contribution vanishes at this order because
\be\label{ghost0}
\frac{\pa}{\pa u^\l} \Tr \big(\c_\l \c^{\r\s} S_0(v-u)\big)  \;\propto \;
     \d(u-v) \, \Tr \c^{\r\s} \,=\, 0
\ee
(but the ghost contribution does {\em not} necessarily vanish for the other gauge choices).

At second order we have two contributions. The first one arises from the 
application of $d/dg$ to (\ref{bR1}) 
\be\label{S2a}
- \, \frac1{2r_D} \int du\, dv\, dw\; \Pi_{\mu\nu} (x-u)\Tr \Big( \c_\nu \c^{\r\s} S^{cm}(v-w) f^{mnp} \Asl^p(w) 
     S^{na}(w-u) \Big)  f^{cde} A_\r^d(v) A_\s^e(v)
\ee
The ghost contribution contained in this expression simplifies to
\be\label{ghost1}
+\,  \frac1{2r_D} \int du\,dv\,dw\; \pa_\mu C(x-u) \Tr \left( \c_\l \c^{\r\s} S^{cm}(v-w) 
     f^{mnp} \Asl^p(w) \frac{\pa S^{na}(w-u)}{\pa u^\l} \right)  f^{cde} A_\r^d(v) A_\s^e(v)
\ee
Working out the gamma traces, with
\be
\frac1{2r_D} \Tr \big(\c_\a \c_\l \c_\b \c_\mu \c^{\r\s} \big) = 
- \d_{\a\l} \d^{\r\s}_{\b\mu}   +  \d_{\a\b} \d^{\r\s}_{\l\mu}   - \d_{\a\mu} \d^{\r\s}_{\l\b} 
 - \d_{\b\mu} \d^{\r\s}_{\a\l}  + \d_{\l\mu} \d^{\r\s}_{\a\b} - \d_{\l\b} \d^{\r\s}_{\a\mu} 
\ee
setting $g=0$ (so $S^{ab}$ is again replaced by $\d^{ab} S_0$), and expressing everything in terms of the scalar propagator, the non-ghost part gives
\begin{align}
f^{abc} f^{bde}
    \int dv\, dw\; 
    \Big[ & -\pa_\r C(x-v) A_\s^c(v) \pa_\s C(v-w) A_\r^d(w) A_\mu^e(w) \nl
& + \pa_\r C(x-v) A_\s^c(v) \pa_\r C(v-w) A_\s^d(w) A_\mu^e(w) \nl[1mm]
& -\pa_\r C(x-v) A_\s^c(v) \pa_\mu C(v-w) A_\s^d(w) A_\r^e(w) \nl[1mm]
& -\pa_\mu C(x-v) A_\r^c(v) \pa_\s C(v-w) A_\s^d(w) A_\r^e(w) \nl[1mm]
& + \pa_\r C(x-v) A_\mu^c(v) \pa_\s C(v-w) A_\s^d(w) A_\r^e(w) \nl[1mm]
& -\pa_\r C(x-v) A_\r^c(v) \pa_\s C(v-w) A_\s^d(w) A_\mu^e(w) \Big]
\end{align}
while (\ref{ghost1}) reduces to
\be
- \, f^{abc} f^{bde} \int dv\,dw\; \pa_\mu C(x-v) A_\r^c(v) \pa_\s C(v-w) A_\r^d(w) A_\s^e(w)
\ee
We thus see that the gauge transformation term $\propto \, \pa_\mu(\cdots)$ cancels between the two expressions: the effect of the transversal projection is precisely to remove any longitudinal terms. 

A second set of terms comes from applying $\bfR$ 
to the $A_\rho^e (u) A_\sigma^d(u)$ in (\ref{bR1}) which gives
\bea\label{bR2}
&+&  \frac1{2r_D^2} \int du\,dv\,dw\,dz\; 
\Pi_{\mu\nu}(x-u)\, {\rm Tr}\, \big( \c_\nu \c^{\r\s} S^{ba}(v-u)\big) f^{bcd} A_\r^c(v) \,
\times \nl[2mm]
&& \quad\quad \times \; \Pi_{\s\t}(v-w) \, \Tr \big(\c_\t \c^{\a\b} S^{ed}(z-w)\big) 
f^{efg} A_\a^f(z) A_\b^g(z)
\eea
As before we set $g=0$ (the ghost term does not contribute, again because of (\ref{ghost0})) 
to get
\begin{align}
f^{abc} f^{bde}
    \int dv\,dw\; &
    \Big[  - \pa_\r C(x-v) A_\mu^c(v) \pa_\s C(v-w) A_\r^d(w) A_\s^e(w) \nl
& + \pa_\r C(x-v) A_\r^c(v) \pa_\s C(v-w) A_\mu^d(w) A_\s^e(w) \Big]
\end{align}
Adding up all the contributions (and the factor $\frac12$ from the Taylor expansion) we obtain
\begin{align}\label{Tg2inv}
(T_g^{-1} A)^a_\mu(x) \,&=\,  A_\mu^a(x) \, - \, 
g f^{abc} \int du\, \pa_\l C(x-u) A_\mu^b(u) A_\l^c(u)  \nl[1mm]
& \; + \,\frac12 g^2 f^{abc} f^{bde} \int dv dw\, 
    \Big[  -\pa_\r C(x-v) A_\s^c(v) \pa_\s C(v-w) A_\r^d(w) A_\mu^e(w) \nl
&  \qquad\qquad+ \pa_\r C(x-v) A_\s^c(v) \pa_\r C(v-w) A_\s^d(w) A_\mu^e(w) \nl[1mm]
& \qquad\qquad   -\pa_\r C(x-v) A_\s^c(v) \pa_\mu C(v-w) A_\s^d(w) A_\r^e(w) \nl[1mm]
& \qquad\qquad + 2 \,\pa_\r C(x-v) A_\mu^c(v) \pa_\s C(v-w) A_\s^d(w) A_\r^e(w) \nl[1mm]
& \qquad\qquad  -2 \, \pa_\r C(x-v) A_\r^c(v) \pa_\s C(v-w) A_\s^d(w) A_\mu^e(w) \Big]
  \; + \;\cO(g^3)
\end{align}
Inverting this result we obtain the map up to second order\footnote{Of course, with 
$3 f^{bde} \pa_{[\mu} C A_\l^d A_{\r ]}^e \,\equiv\,  f^{bde} \big(\pa_{\mu} C A_\l^d A_{\r }^e \,+\,
\pa_{\l} C A_\r^d A_{\mu }^e \,+\, \pa_{\r} C A_\mu^d A_{\l }^e\big) $.}
\bea\label{Tg2}
(T_gA)^a_\mu(x) \,&=&\,  A_\mu^a(x) \, + \, g f^{abc} \int du\, \pa_\l C(x-u) A_\mu^b(u) A_\l^c(u)
 \\[2mm]
 &&  + \, \frac32 \, g^2f^{abc}f^{bde} \int du dv \, \pa_\r C(x-u) A_\l^c(u)
        \pa_{[\mu} C(u-v) A_\l^d (v) A_{\r]}^e(v)  \, + \, \cO(g^3) \nonumber
\eea
which agrees with the original result (Equation (3.24) from \cite{Nic1}). As mentioned already, both (\ref{Tg2inv}) and (\ref{Tg2})  can be read with both Euclidean and Minkowskian signatures, respectively. For some simple (but non-trivial) quantum correlators involving scalar operators of the $N=4$ theory these formulas do give results which, up to $\cO(g^2)$, precisely agree with those obtained using more standard techniques \cite{NP}. These computations also confirm the claim of \cite{DL1,DL2} that the amount of labor required to determine quantum correlators by means of this ghost and fermion free formalism is comparable to the usual one.

\section{Jacobians, fermion and ghost determinants to $\cO(g^2)$}

In this section, we check the main statement. First, it is easily verified that
\be
\pa^\mu A^{'a}_\mu(x) \,=\, \pa^\mu A_\mu^a(x) + \cO(g^3)
\ee
Likewise, a straightforward calculation shows that
\be
\frac12  \int d^D \, x\, \Big[ A^{'a}_\mu (-\Box) A^{'a}_\mu \,-\, (\partial^\mu A^{'a}_\mu))^2\Big] \,=\, 
    \frac14 \int d^D x\,  F_{\mu\nu}^a F_{\mu\nu}^a + \cO(g^3)
\ee
with the $g$-dependent Yang-Mills field strength (\ref{Fmn}) on the r.h.s.
These parts of the calculation do {\em not} make use of the special value of $D$, and therefore work in all dimensions. 

The dependence on the dimension enters only through the second part (\ref{Det}). 
For the (perturbative) computation of the relevant functional determinants (or rather their logarithms)  we use the standard formula
\be
\log \det \big(\bE-\bX\big) = \Tr \log \big(\bE - \bX \big) = - \sum_{n=1}^\infty \frac1{n} 
{\rm Tr}\, \bX^n
\ee 

Let us first consider the Jacobian corresponding to (\ref {Tg2}). To first order it simply vanishes because $f^{aac} =0$ (or alternatively, $\pa_\l C(0)=0$). After a little computation we arrive at the following result 
\bea\label{Jacdet}
\log \det \left(\frac{\delta A^{'a}_\mu(x,g;A)}{\delta A^b_\nu(y)}\right) &=&
\frac12 n g^2 \int dx\,dy\;\Big\{ (2D-3) \pa_\mu C(x-y) A_\mu^a(y) \pa_\nu C(y-x) A_\nu^a(x) \nn\\
&& \hspace{-0.2cm}  - (D-2) \pa_\mu C(x-y) A_\nu^a(y) \pa_\mu C(y-x) A_\nu^a(x) \Big\}
+ \cO(g^3)
\eea
where we have used $f^{gcd} f^{hcd} = n\, \delta^{gh}$ and the relation
\be
\int dx\,dy\;\pa_\mu C(x-y) A_\mu^a(y) \pa_\nu C(y-x) A_\nu^a(x)  =
\int dx\,dy\; \pa_\mu C(x-y) A_\nu^a(y) \pa_\nu C(y-x) A_\mu^a(x) 
\ee
which follows by partial integration and use of the Landau gauge condition $\pa^\mu A_\mu^a = 0$ (or, more precisely, the presence of the $\delta$-functional in the functional measure (\ref{deltafunctional})).

For the ghost determinant the relevant functional matrix is
\be
\bX^{ab}(x,y;A) \,=\, g f^{abm} C(x-y) A_\mu^m(y) \pa_\mu^y
\ee
which gives
\be
\log \det \big(\bE - \bX\big) = 
\frac12 \, n g^2 
\int dx\,dy\; \pa_\mu C(x-y) A^m_\nu(y) \pa_\nu C(y-x) A_\mu^n(x)
     \,+ \, \cO(g^3)
\ee
Observe that the $\cO(g)$ term vanishes as before. The $\cO(g^2)$ term has again been simplified by using $f^{abc} f^{bad} = - n\,\d^{cd}$. Because $\det (\pa^\mu D_\mu) = \det (D_\mu \pa^\mu)$, we can equivalently write
\bea
\log \det \big(\bE - \bX\big) &=& 
\frac12 \, n g^2 
\int dx\,dy\; \pa_\mu C(x-y) A^m_\mu(y) \pa_\nu C(y-x) A_\nu^n(x)
    \,+\,  \cO(g^3)
\eea
an equality that can also be checked explicitly by partial integration and use of $\pa^\mu A_\mu^a = 0$.

For the Matthews-Salam determinant we have (suppressing spinor indices)
\be
\bY^{ab}(x,y;A) \,=\,  g f^{abm} \pa_\a C(x-y) \c^\a \c^\l A_\l^m(y)
\ee
With an extra overall factor of $\frac12$ for Majorana fermions we get
\bea
\frac12 \log \det \big(\bE - \bY\big) &=& \frac14  n g^2  
\Tr (\c_\a \c_\l \c_\b \c_\r ) \int dx\,dy\;
      \pa_\a C(x-y) A_\l^m(y) \pa_\b C(y-x) A_\r^m (x) \nn\\
  && + \;\; \cO(g^3) 
\eea
Adding all the terms and demanding equality with (\ref{Jacdet}) yields two conditions
 \be
 2D-3 = 1+r_D \;\;, \quad D-2 = \frac{r_D}2
 \ee
 which happily coincide and are thus satisfied for 
\be
D\,=\,  3,4,6,10  \qquad \Longleftrightarrow \qquad r_D\,=\, 2,4,8,16
\ee
(but not for any other values of $D$ and $r_D\,$). We therefore recover the old result of \cite{SYM0} without any use of anticommuting objects whatsoever.\footnote{For the {\em free} theory this
equality follows trivially by demanding cancellation of the free determinants, with
$$
\int \cD A \, e^{\frac12 A\Box A} \,\sim\, [\det(-\Box)]^{-D/2} \;\; , \quad
\int \cD C \cD \bar{C} \, e^{\bar C \Box C} \,\sim\, \det (-\Box)\;\; , \quad
\int \cD\chi \, e^{\frac12 \bar\chi \pasl\chi} \,\sim \, [(\det(-\Box)]^{r_D/4}
$$
which is just the statement that bosonic and fermionic degrees of freedom must match on shell.}
Given that our statement about the permitted dimensions applies to the interacting theory, it is 
effectively equivalent to the more standard calculation to verify the closure of supersymmetry 
transformations which requires the use of a specific Fierz identity that is valid only 
for $D\,=\,  3,4,6,10$ \cite{SYM1}.

At higher orders, the calculations presented here become technically involved fairly quickly.  While the procedure to derive the inverse map is rigorous it does prove lengthy, with $\cO(n!)$ terms at order $\cO(g^n)$. It will thus be interesting to see whether there exists an algorithmic approach that leads directly to the map itself~\cite{SYM2} (see \cite{L2} for earlier work in this direction). The existence of a simpler algorithm for $T_g$ is also suggested by the comparison of formulas (\ref{Tg2}) and (\ref{Tg2inv}), and the fact that the MSS and the FP determinants $\Delta_{MSS}$ and $\Delta_{FP}$ involve only structures of the type $\pa C A \cdots \pa C A$, whereas the $\cR$-prescription leads to various other structures as well. All this indicates that the map $T_g$ itself may have a simpler structure than the inverse map $T_g^{-1}$, a feature that can also be seen in other examples.

\section{Afterword: personal memories of Peter Freund (by H.N.)}

{\it This final section contains some of H.N.'s personal reminiscences of his encounters 
with Peter Freund since the early 80ies.}

Peter Freund was a source of numerous unusual and fertile ideas in physics that
also inspired parts of my own early work, and thus had an important 
influence on my career. Perhaps best remembered is the pivotal role he played in 
the development of modern Kaluza-Klein theories, thus contributing to their revival
after many decades of dormancy \cite{CF,FR}.

I had the privilege of meeting Peter Freund many times, in particular on the occasion
of several visits to Chicago. But in the early 80ies he also came to CERN, where I was 
employed as a junior staff member of the CERN Theory Division at the time. On one of these visits
(as far as I remember, in  the wake of my work with Bernard de Wit on $N=8$ supergravity)
he enquired whether I would be interested in joining the University of Chicago as a junior
faculty member (on what I suppose is nowadays called a tenure track position). 
For me this was definitely a very attractive option, 
but I finally did not move to the US, mainly for family reasons, settling for a less glamorous 
position at the University of Karlsruhe (with Julius Wess).

Our main scientific overlap in those days was Kaluza-Klein supergravity,
which centered to a large extent on the famous Freund-Rubin solution \cite{FR}, the 
first real and concrete example of a theory with {\em preferential} compactification to 
four dimensions -- and still the only one, as far as I am aware!
In fact, at about the same time, with Antonio Aurilia and Paul Townsend we had also been
wondering about the meaning of an expectation value for the 4-form field \cite{ANT}, 
showing that the cosmological constant could be interpreted as an integration constant
(and hence its value somehow be endowed with a dynamical origin). 
Regrettably, however, as Paul Townsend aptly put it later, we did ``miss the boat on 
the Freund-Rubin solution"!

About two years later the heterotic string \cite{GHMR} appeared on stage, eclipsing
everything else that had come before, and rolling over the  CERN Theory Division
like a tsunami. Offering for the first time real prospects for linking string theory to Standard
Model physics there immediately arose the question what the link was between
this totally new theory and the more established purely bosonic string or the superstring. 
It was again Peter Freund who (after early premonitions of E$_8\,\times\,$E$_8$) stepped 
forward with an audacious idea, namely the proposal that the heterotic string was 
actually some compactified version (though of a strange kind) of the bosonic string 
in $D=26$ \cite{Het}. This idea crucially inspired our own work \cite{CENT}. I was actually
amazed at all the attention we got for that paper -- for a while this was the only thing 
people wanted to hear about from me! I even got an invitation from Murray Gell-Mann to his 
newly founded Santa Fe Institute to speak about this work, a task that I accepted with 
considerable trepidation because I was aware that the select audience there would  
consist of some of the smartest minds on the planet, some of whom I knew were
not particularly positively inclined towards the idea 
(I still remember David Gross greeting me at the local airport with the words 
``We have seen your paper, and we don't believe your claims"). Looking back, it must be said 
that the idea finally did not fly as we had hoped, remaining mostly a kinematic scheme,
and has by now largely faded away from the string landscape (like so many other ideas).

An especially memorable encounter happened in 2008 when Peter invited me to Timisoara 
to deliver the annual Schr\"odinger Lecture at the local university, 
an event sponsored by the University of Vienna, of which Peter had been
put in charge in recognition of his enduring attachment to the old world cultural 
charm of the no longer existing Austro-Hungarian empire. Timisoara is the place
where Peter was born. On the occasion of this visit he showed me many 
of the places of his early childhood, telling me about his multicultural upbringing
and how he grew up learning to speak so many different
languages (Romanian, Hungarian and German, for starters), that later enabled him
to become such an impressive polyglot. And he also told me how as a child he only barely
escaped the Nazi terror, largely crediting the ineffectiveness of the Romanian bureaucracy 
for saving his life, because these people had been neither eager nor efficient in 
implementing the invaders' new rules.

On my last visit to Chicago, he invited me not only to a performance of
the Chicago Symphony Orchestra (including a Sibelius symphony which we both
found boring), but took me along to some fancy reception at the local Austrian 
Consulate on top of the Lake Point Tower building, an architechtural landmark
right on the shore of Lake Michigan, to 
which he had been invited for some reason. I had obviously {\em not} been invited, but 
went along anyway, though with a bit of embarrassment as our dress code did
not match the standards expected at such an event.  Nevertheless, up on the 67th floor, 
and without being taken much note of by the Austrian Consul nor his other guests, we 
had a great time, enjoying the food and the wine, with fabulous views 
of Lake Michigan and the surrounding Chicago skyline.

The last time I saw him was on occasion of his visit to Berlin and to AEI in Potsdam
where he delivered a colloquium on (also his!) ``passion for physics", and I had the 
the opportunity to invite him (and Jan Plefka) to dinner in the rooftop restaurant on top
of the Reichstag, with a splendid vista of the Berlin night sky which he enjoyed very much. 
But the thing that sticks in my mind more than anything else is Peter's great love and
appreciation of music, especially of the vocal kind. So most of our `off-physics talk'
revolved around music, with me learning a lot about his preferences, and also his `dislikes'
in musical matters (for instance, he admired Richard Wagner and Francis Poulenc, 
but had no appreciation at all for Bach Cantatas). Not just being a musical expert, he  was 
also a quasi-professional singer and performer (as people in Chicago will surely remember). 
This is something you would also notice when he gave physics seminars, which always
had a kind of operatic touch (and I often thought of Don Giovanni singing on stage
while listening to him).  Accordingly, my visits to all of his three places in Chicago 
would invariably end with us trying (with me on the piano) to do some of the highlights 
of the Lieder repertoire, such as {\em Die Winterreise} by Schubert, or various Schumann 
Lieder (some of which, by the way,  he treasured  as the absolute culmination point of 
this musical genre). 

I will fondly remember Peter Freund as a great friend and a great 
physicist.
\vskip 0.5cm
\noindent {\bf Acknowledgments:} H.N. would like to thank O. Lechtenfeld  and J. Plefka
for discussions and correspondence, and IISER Pune for hospitality in November 2019.

\end{document}